\documentclass[a4paper,11pt]{article} 
\usepackage[margin=3cm]{geometry}
\usepackage{amstext,amsmath,amssymb}
\usepackage{blindtext}
\usepackage{enumitem}
\usepackage{changepage, environ}
\PassOptionsToPackage{pass,showframe}{ge	ometry}
\usepackage{float}
\usepackage[utf8]{inputenc}
\usepackage[T1]{fontenc}
\usepackage[autostyle]{csquotes}
\usepackage{dirtytalk}
\usepackage{graphicx} 
\usepackage{caption}
\usepackage[english]{babel}
\usepackage{xcolor}
\usepackage{breakurl}
\usepackage[breaklinks]{hyperref}
\usepackage{setspace}
\hypersetup{colorlinks=true, pdfstartview=FitV, linkcolor=blue, citecolor=black, plainpages=false, pdfpagelabels=true, urlcolor=blue}
\usepackage{xurl}
\usepackage{scalerel}
\usepackage{tikz}
\usepackage{multicol}
\usepackage{blindtext}
\usepackage{abstract}
\usepackage{sectsty}
\usepackage{secdot}
\usepackage{fancyhdr}
\usepackage{secdot}
\usepackage{blindtext}
\usepackage{academicons}
\usepackage{graphicx}
\graphicspath{ {./images/} }

\PassOptionsToPackage{dvipsnames}{xcolor}
\usetikzlibrary{decorations.text}
\usepackage{tikz}
\usepackage[nottoc,numbib]{tocbibind}
\usepackage{setspace}

\IfFileExists{upquote.sty}{\usepackage{upquote}}{}
\IfFileExists{microtype.sty}{
  \usepackage[]{microtype}
  \UseMicrotypeSet[protrusion]{basicmath} 
}{}
\makeatletter
\@ifundefined{KOMAClassName}{
  \IfFileExists{parskip.sty}{%
    \usepackage{parskip}
  }{
    \setlength{\parindent}{0pt}
    \setlength{\parskip}{6pt plus 2pt minus 1pt}}
}{
  \KOMAoptions{parskip=half}}
\makeatother
\usepackage{longtable,booktabs,array}
\usepackage{calc} 

\title{\textbf{Affective Computing and Emotional Data: Challenges and Implications in Privacy Regulations, The AI Act, and Ethics in Large Language Models}}

\usepackage[multiple]{footmisc}
\usepackage{bigfoot}

\DeclareNewFootnote{AAffil}[arabic]
\DeclareNewFootnote{ANote}[fnsymbol]

\NewEnviron{myquote}[1][\relax]{
  \begin{center}
  \begin{adjustwidth}{50pt}{50pt}
  \setbox0=\hbox{\itshape\BODY}%
  \begin{itshape}
  \ifdim\wd0>\linewidth\relax\BODY\else\hfil\BODY\fi
  \end{itshape}
  \end{adjustwidth}
  \end{center}  
  \ifx\relax#1\relax\else
    \vspace{-0.15cm}
    \begin{adjustwidth}{0pt}{20pt}
    \flushright { \small -\hspace{2pt}#1\hspace{2pt}- }
    \end{adjustwidth}
    \vspace{0.3cm}
  \fi
}

\author{
  Nicola Fabiano\thanks{\href{https://www.fabiano.law/en/page/about/}{Studio Legale Fabiano (Italy)} - Affiliation: \textit{International Institute of Informatics and Systemics (IIIS) - USA}}
}

\date{August 29, 2025}

\begin{document}
\maketitle

\begin{abstract}
This paper examines the integration of emotional intelligence into artificial intelligence systems, with a focus on affective computing and the growing capabilities of Large Language Models (LLMs), such as ChatGPT and Claude, to recognize and respond to human emotions. Drawing on interdisciplinary research that combines computer science, psychology, and neuroscience, the study analyzes foundational neural architectures—CNNs for processing facial expressions and RNNs for sequential data, such as speech and text—that enable emotion recognition. It examines the transformation of human emotional experiences into structured emotional data, addressing the distinction between explicit emotional data collected with informed consent in research settings and implicit data gathered passively through everyday digital interactions. That raises critical concerns about lawful processing, AI transparency, and individual autonomy over emotional expressions in digital environments. The paper explores implications across various domains, including healthcare, education, and customer service, while addressing challenges of cultural variations in emotional expression and potential biases in emotion recognition systems across different demographic groups. From a regulatory perspective, the paper examines emotional data in the context of the GDPR and the EU AI Act frameworks, highlighting how emotional data may be considered sensitive personal data that requires robust safeguards, including purpose limitation, data minimization, and meaningful consent mechanisms. The case study of OpenAI's ChatGPT-4.5, which incorporates improved emotional intelligence capabilities, exemplifies advances in emotionally responsive AI while illustrating the complexity of ensuring the ethical deployment of such systems. The paper concludes by proposing a multidimensional governance framework that integrates technical scrutiny with legal and ethical analysis, emphasizing multi-stakeholder collaboration to safeguard human dignity, autonomy, and rights in the digital age of emotionally aware AI systems.
\end{abstract}

\noindent \textbf{Keywords.} Artificial Intelligence - Emotional Data - Large Language Models (LLM) - Data Protection - Privacy - Ethics\\

\newpage
\begin{spacing}{0.2}
\tableofcontents
\end{spacing}
\newpage

\section{Introduction}\label{introduction}

The evolution of artificial intelligence (AI) has witnessed a significant shift from systems designed primarily for logical reasoning and task execution to those capable of recognizing, interpreting, and responding to human emotions. This transformation represents a significant advancement in human-computer interaction, as it addresses the emotional dimension of communication that has been traditionally absent from digital interfaces. The field of affective computing, first conceptualized by Rosalind Picard in the late 1990s, has now emerged as a cornerstone of modern AI development, particularly in the realm of Large Language Models (LLMs) such as ChatGPT, Claude, and other conversational agents (Picard, 1997).

Integrating emotional intelligence into AI systems carries profound implications for various domains, including healthcare, education, customer service, and personal assistance. By recognizing emotional states, AI systems can provide more personalized, empathetic, and contextually appropriate responses, potentially enhancing user satisfaction and overall effectiveness. However, this capability also introduces complex challenges related to privacy, consent, data protection, and the fundamental nature of human-AI relationships.

This paper examines the technological underpinnings, regulatory considerations, and ethical implications of emotionally intelligent AI systems. We begin by examining the neural network architectures that enable machines to process emotional cues, followed by an analysis of how human emotions are transformed into structured data. Subsequently, we investigate the regulatory frameworks governing emotional data, with a focus on the European Union's General Data Protection Regulation (GDPR) and the EU Artificial Intelligence Act. Using the case study of OpenAI's ChatGPT-4.5, we illustrate the practical implications of these advancements. Ultimately, we propose a multidimensional governance approach that strikes a balance between technological innovation and the protection of human dignity and autonomy.

\section{Affective Computing: Technical Foundations and Evolution}\label{affective-computing-technical-foundations-and-evolution}

\subsubsection{The Evolution of Affective
Computing}\label{the-evolution-of-affective-computing}

Affective computing emerged as a distinct field of research in 1997, when MIT Media Lab professor Rosalind Picard published her seminal work, "Affective Computing," which defined the discipline as "computing that relates to, arises from, or deliberately influences emotions" (Picard, 1997). This multidisciplinary domain integrates concepts from computer science, psychology, neuroscience, and cognitive science to create systems that can detect, understand, and respond to human emotions.

The evolution of affective computing can be traced through several key developmental phases. Initial efforts in the 1990s and early 2000s focused on identifying basic emotions from facial expressions and vocal patterns using rule-based systems and simple machine learning algorithms. These systems typically operate with predefined emotional categories based on Paul Ekman's six basic emotions: happiness, sadness, anger, fear, disgust, and surprise (Ekman, 1992). As research progressed into the 2000s and 2010s, scientists recognized that human emotional expression involves multiple channels. This led to the development of systems that combined facial, vocal, textual, and physiological data for more robust emotion recognition. This period witnessed the creation of multimodal datasets, such as IEMOCAP ("Interactive Emotional Dyadic Motion Capture" - Busso et al., 2008) and SEMAINE ("Sustained Emotionally colored Machine-human Interaction using Nonverbal Expression" - McKeown et al., 2012), which facilitated advances in multimodal emotion recognition.

The advent of deep learning techniques in the 2010s transformed affective computing, enabling more accurate recognition of subtle emotional cues and complex emotional states. Convolutional Neural Networks (CNNs) transformed facial expression analysis, whereas Recurrent Neural Networks (RNNs), followed by Transformer architectures, brought substantial advancements in recognizing emotions from text and speech. From 2015 onward, research shifted toward context-aware systems that consider situational factors, individual differences, cultural variations, and interaction history when interpreting emotional signals and generating responses. Most recently, from 2020 to the present, the emergence of sophisticated LLMs has accelerated the development of emotionally intelligent AI systems that can engage in nuanced emotional conversations, recognize implicit emotional cues in text, and generate contextually appropriate emotional responses.

\subsubsection{Neural Network Architectures for Emotion Recognition}\label{neural-network-architectures-for-emotion-recognition}

The ability of AI systems to recognize and process emotional data relies primarily on two neural network architectures: Convolutional Neural Networks (CNNs) and Recurrent Neural Networks (RNNs), as well as more recent Transformer-based models. These architectures serve distinct but complementary functions in analyzing different emotional cues.

\subsubsection{Convolutional Neural Networks (CNNs)}\label{convolutional-neural-networks-cnns}

CNNs have revolutionized the field of computer vision through their hierarchical structure, which enables the extraction of increasingly
complex features from visual data (LeCun et al., 2015). In the context of emotion recognition, CNNs are particularly effective for processing
facial expressions, a primary channel of emotional communication among humans. These networks apply convolutional filters across image inputs, detecting patterns such as edge orientations, textures, and higher-level features corresponding to specific emotional expressions.

The effectiveness of CNNs in emotion recognition is exemplified by models such as FER2013 (Facial Expression Recognition) and EmotioNet, which achieve accuracy rates exceeding 70\% in classifying basic emotions from facial images (Goodfellow et al., 2013; Fabian Benitez-Quiroz et al., 2016). Moreover, advanced architectures like ResNet and EfficientNet have enhanced performance through deeper networks with residual connections and optimized scaling methods (He et al., 2016; Tan \& Le, 2019).

The application of CNNs extends beyond static image analysis to dynamic facial expressions in video streams, where temporal convolutional networks (TCNs) can capture the evolution of expressions over time. This capability is crucial for detecting subtle emotional transitions and distinguishing between genuine and feigned expressions, a distinction that often depends on temporal dynamics (Ekman \& Friesen, 1982).

\subsubsection{Recurrent Neural Networks (RNNs) and Transformers}\label{recurrent-neural-networks-rnns-and-transformers}

While CNNs excel at processing spatial data, RNNs are designed to handle sequential information, making them ideal for analyzing temporal aspects of emotional expression in speech, text, and physiological signals (Hochreiter \& Schmidhuber, 1997). The distinctive feature of RNNs is their internal memory, which enables them to maintain context across sequential inputs — a characteristic essential for understanding emotions that unfold and evolve.

Variants of RNNs, including Long Short-Term Memory (LSTM) networks and Gated Recurrent Units (GRUs), have demonstrated remarkable success in sentiment analysis and emotion recognition from textual and audio data.
For instance, DeepMoji, an LSTM-based model trained on a dataset of 1.2 billion tweets containing emojis, achieves state-of-the-art performance in detecting emotions and sentiment in text (Felbo et al., 2017).

Recently, Transformer architectures have surpassed traditional RNNs in many natural language processing tasks, including emotion recognition. Models such as BERT (Bidirectional Encoder Representations from Transformers) and its successors leverage attention mechanisms to capture subtle emotional nuances in text by dynamically assessing the contextual relevance of words (Devlin et al., 2019). This ability is beneficial for interpreting complex emotional states that are conveyed implicitly, rather than through direct emotional language.

The latest generation of LLMs, including GPT-4 and Claude, builds upon these Transformer architectures, incorporating vast amounts of textual data that enable them to recognize and respond to emotional cues with unprecedented sophistication. These models can detect basic emotions and complex emotional states, as well as cultural variations in emotional expression and contextual factors that influence emotional interpretation.

\subsubsection{Multimodal Approaches to Emotion Recognition}\label{multimodal-approaches-to-emotion-recognition}

Human emotional expression is inherently multimodal, involving a combination of facial expressions, vocal intonations, linguistic choices, body language, and physiological responses. Recognizing this complexity, contemporary AI systems are increasingly adopting multimodal approaches that integrate information from multiple channels to achieve more robust and accurate emotion recognition.

Multimodal fusion techniques combine data from different modalities at various processing levels. Early fusion approaches concatenate raw data from other modalities before feature extraction, allowing the model to learn joint representations directly from the multimodal input.
Late fusion methods process each modality independently through separate models, and their predictions are combined at the decision level through weighted averaging or voting methods. Hybrid fusion techniques combine early and late fusion elements, with intermediate features from different modalities being integrated at multiple levels throughout the processing pipeline.

Research has consistently demonstrated that multimodal approaches outperform unimodal methods in emotion recognition tasks. For example, the IEMOCAP (Interactive Emotional Dyadic Motion Capture) dataset, which includes audio, visual, and textual information from acted emotional scenes, has been used to develop models that achieve accuracy improvements of up to 10\% when using multimodal versus unimodal approaches (Busso et al., 2008).

Recent advancements in multimodal learning include cross-modal attention mechanisms, which allow models to selectively focus on relevant information across modalities, and contrastive learning techniques that align representations from different modalities in a shared embedding space. These methods are particularly valuable for handling the inherent asynchrony and complementarity of emotional cues across various channels.

\section{Emotional Data: Conceptualization, Collection, and Processing}\label{emotional-data-conceptualization-collection-and-processing}

\subsubsection{Conceptualizing Emotional Data}\label{conceptualizing-emotional-data}

The concept of ``emotional data'' represents a significant departure from traditional data types in computing. Unlike factual information, emotional data encompasses subjective, experiential, and often ambiguous aspects of human emotional states. This transformation — from lived emotional experiences to structured, machine-processable data — raises fundamental questions about the nature of emotions and the extent to which they can be captured and represented digitally.

Emotional data can be categorized in several ways. First, we can distinguish between explicit and implicit emotional data. Explicit emotional data includes direct self-reports of emotional states, such as survey responses or emotion labels provided by users. Implicit emotional data, in contrast, is derived from behavioral signals that may indicate emotional states, such as facial expressions, voice patterns, or physiological responses. Second, emotional data can be approached either categorically or dimensionally. Categorical approaches classify emotions discretely (e.g., joy, sadness, anger).
In contrast, dimensional approaches represent emotions along continuous axes such as valence (positive/negative), arousal (high/low energy), and dominance (high/low control). Third, emotional data may be individual or aggregate. Individual emotional data pertains to specific persons and their unique emotional experiences, whereas aggregate emotional data represents patterns across populations or groups. Finally, emotional data can be context-dependent or context-independent. Context-dependent emotional data incorporates situational factors that influence emotional experiences, while context-independent data focuses solely on the emotional signals.

The transformation of emotions into data necessarily involves simplification and abstraction, potentially obscuring the richness and complexity of human emotional experience. As Barrett (2017) noted, emotions are not singular, universal phenomena but constructed experiences influenced by cultural, social, and individual factors. This perspective challenges the notion that emotions can be reduced to discrete, objective data points without significant loss of meaning.

\subsubsection{Methods of Emotional Data Collection}\label{methods-of-emotional-data-collection}

The collection of emotional data employs various methodologies, each with distinct strengths, limitations, and ethical implications. Direct self-report methods include questionnaires, experience sampling methods (ESM), and ecological momentary assessment (EMA), in which individuals directly report their emotional states. While these methods provide valuable insights into subjective experiences, they are subject to limitations such as social desirability bias, recall errors, and variations in emotional vocabulary and awareness across individuals (Robinson \& Clore, 2002). Digital implementations of self-report measures, such as mood-tracking applications and emotion rating systems, facilitate the continuous collection of emotional data in naturalistic settings. However, the frequency and intrusiveness of such measures must be balanced against user burden and potential reactivity effects, where reporting emotions influences the emotional experience itself.

Observational approaches capture behavioral manifestations of emotions without requiring explicit reporting from individuals. These methods encompass facial expression analysis, utilizing computer vision techniques to detect and classify facial expressions based on the Facial Action Coding System (FACS) or deep learning approaches. They also include voice analysis, extracting acoustic features such as pitch, tempo, and spectral characteristics to identify emotional cues in speech. Text analysis employs natural language processing techniques to identify emotional content in written communications, encompassing sentiment analysis, emotion classification, and stance detection. Physiological measurements monitor autonomic nervous system responses such as heart rate variability, electrodermal activity, and respiratory patterns, which correlate with emotional states. Additionally, behavioral metrics track user interactions with digital systems, including click patterns, dwell times, and navigation behaviors, which may reflect emotional engagement or frustration.

The context in which emotional data is collected significantly influences both the nature of the data and the ethical considerations surrounding its use. Emotional data is typically collected in controlled research environments, where explicit informed consent, clear protocols, and institutional oversight are in place. Participants are generally aware of the specific data and how it will be used. Consumer-facing products, however, may collect emotional data through various channels, sometimes without users' full awareness or understanding. This data may be used for personalization, engagement optimization, or marketing. Ambient intelligence environments, such as smart homes, workplaces, and public spaces, increasingly incorporate sensors that can passively collect data indicative of emotional states, often without requiring active user participation in the data collection process. Social media sites, forums, and other digital platforms generate vast repositories of emotional data through user-generated content, engagement metrics, and interaction patterns.

The distinction between explicit and implicit collection contexts has profound implications for consent, transparency, and user autonomy.
While users in research settings typically provide informed consent for specific data collection activities, users of commercial products or public spaces may have limited awareness of emotional data collection and few opportunities to consent or opt out meaningfully.

\subsubsection{Representation and Processing of Emotional Data}\label{representation-and-processing-of-emotional-data}

Once collected, emotional data undergoes various transformations to facilitate machine processing and analysis. Emotional data may be represented in multiple formats depending on the source and intended application. Feature vectors provide numerical representations of extracted features, such as facial landmark coordinates, acoustic parameters, or linguistic features. Categorical labels classify emotions into predefined categories (e.g., Ekman's six basic emotions: happiness, sadness, fear, disgust, anger, and surprise).
Dimensional coordinates position emotional states in multidimensional spaces, most commonly the valence-arousal-dominance (VAD) model.
Time-series data captures the dynamics of emotional expression over time, including onset, peak, and decay patterns. Probability distributions represent uncertainty or ambiguity in emotional classification, acknowledging that emotional states often involve blends or transitions between prototypical emotions.

The processing of emotional data typically involves several stages.
Preprocessing encompasses cleaning, normalizing, and aligning raw data, including noise reduction, artifact removal, and temporal synchronization across modalities. Feature extraction identifies relevant attributes that convey emotional information, utilizing handcrafted features informed by domain knowledge and learned features derived through representation learning techniques. Fusion integrates information from multiple sources or modalities to create comprehensive emotional profiles. Classification or regression applies machine learning algorithms to categorize or position emotional states along continuous dimensions. Interpretation contextualizes emotional data within broader user profiles, interaction histories, or situational factors to derive meaningful insights. Finally, response generation formulates system responses based on recognized emotional states, potentially incorporating empathetic language, adaptive interface elements, or personalized content.

\section{Regulatory Frameworks for Emotional Data}\label{regulatory-frameworks-for-emotional-data}

\subsubsection{GDPR and Emotional Data Classification}\label{gdpr-and-emotional-data-classification}

The European Union's General Data Protection Regulation (GDPR) provides one of the most comprehensive frameworks for understanding the legal status of emotional data. However, it does not explicitly address this category. Several provisions are particularly relevant to handling emotional data in AI systems.

\subsubsection{Emotional Data as Personal Data}\label{emotional-data-as-personal-data}

The GDPR defines personal data as ``any information relating to an identified or identifiable natural person (‘data subject’)'' (Article 4(1)).
Emotional data falls within this broad definition, as it relates directly to an individual's internal states and can often be linked to specific persons. Consequently, the collection and processing of emotional data must adhere to the general data protection principles established in the regulation.

The principle of lawfulness, fairness, and transparency (Article 5(1)(a)) requires that processing must be based on a legitimate legal ground, conducted fairly, and transparently disclosed to the data subject. For emotional data, this principle demands clear communication about when and how emotional states are being analyzed, a particularly challenging requirement given the often implicit nature of emotional data collection in modern systems.

Purpose limitation (Article 5(1)(b)) stipulates that personal data must be collected for specified, explicit, and legitimate purposes and not further processed in a manner incompatible with those purposes. This principle restricts repurposing emotional data collected for one service (personalization) for unrelated purposes (such as targeted advertising based on emotional vulnerabilities).

Data minimization (Article 5(1)(c)) states that personal data shall be ``\textit{adequate, relevant and limited to what is necessary in relation to the purposes for which they are processed (‘data minimisation’)}''. This principle challenges the common practice of broad emotional data collection that exceeds functional requirements, often justified by potential future applications or system improvements.

Accuracy (Article 5(1)(d)) requires that personal data shall be ``\textit{accurate and, where necessary, kept up to date; every reasonable step must be taken to ensure that personal data that are inaccurate, having regard to the purposes for which they are processed, are erased or rectified without delay (‘accuracy’)}''. This principle is particularly challenging for emotional data, given the inherent subjectivity of emotional interpretation and the potential for misclassification across different demographic groups or cultural contexts.

Storage limitation (Article 5(1)(e)) dictates that personal data should be ``\textit{kept in a form that permits the identification of data subjects for no longer than necessary for the purposes for which the personal data are processed}''. This principle challenges the indefinite retention of emotional profiles, which may become increasingly comprehensive and invasive over time.

Integrity and confidentiality (Article 5(1)(f)) mandate that personal data shall be ``\textit{processed in a manner that ensures appropriate security of the personal data, including protection against unauthorised or unlawful processing and against accidental loss, destruction or damage, using appropriate technical or organisational measures (‘integrity and confidentiality’)}''. This principle acknowledges the sensitive nature of emotional information and the potential harm that can result from unauthorized access or disclosure.

\subsubsection{Emotional Data as Special Category Data}\label{emotional-data-as-special-category-data}

More significantly, emotional data may qualify as ``special category data'' under Article 9 of the GDPR, which includes ``data concerning health'' and potentially ``biometric data to uniquely identify a natural person.'' This classification is particularly relevant when emotional data is derived from physiological measurements that reveal information about mental health conditions, used to infer psychological characteristics or behavioral patterns, collected through biometric methods such as facial expression analysis, or can reveal sensitive aspects of an individual's identity or condition.

If classified as unique category data, emotional data processing would be prohibited unless a specific exception applies, such as explicit consent (Article 9(2)(a)) or when processing is necessary for scientific research purposes, subject to appropriate safeguards (Article 9(2)(j)).
This heightened level of protection recognizes the intimate nature of emotional information and its potential to reveal deeply personal aspects of an individual's life and identity.

\subsection{Legal Basis for Processing Emotional Data}\label{legal-basis-for-processing-emotional-data}

Under the GDPR, any processing of personal data, including emotional data, requires a lawful basis. Several potential bases may apply for emotional data, each with distinct implications for designing and deploying emotionally intelligent systems.

Consent (Article 6(1)(a)) requires that the data subject has given clear, specific, informed, and unambiguous consent to the processing of their emotional data. While seemingly straightforward, obtaining meaningful consent for the processing of emotional data faces significant challenges.
Users may struggle to understand the full implications of emotional analysis, mainly when the technology operates invisibly or implicitly. Furthermore, the power imbalances between individuals and technology providers may undermine the voluntary nature of consent, particularly when emotional analysis is embedded in essential services or platforms.

Legitimate interests (Article 6(1)(f)) states: ``\textit{processing is necessary for the purposes of the legitimate interests pursued by the controller or by a third party, except where such interests are overridden by the interests or fundamental rights and freedoms of the data subject which require protection of personal data, in particular where the data subject is a child}''. This basis requires a careful balancing test that weighs the benefits of emotional analysis against potential harm to individual rights and freedoms. The intimate and potentially revealing nature of emotional data may often tip this balance toward more effective protection of personal interests, especially in commercial applications where the primary legitimate interest is profit-driven rather than welfare-enhancing.

According to Article 6(1)(b), ``processing is necessary for the performance of a contract to which the data subject is party or in order to take steps at the request of the data subject prior to entering into a contract''. That case might be when the user explicitly requests an emotionally responsive service. This basis is most applicable when emotional analysis forms a core function of the service rather than an ancillary feature and when emotional responsiveness is communicated as part of the service offering.

In the case of research purposes (Article 6(1)(e), in conjunction with Article 89), processing is lawful when it is ``\textit{necessary for the performance of a task carried out in the public interest or in the exercise of official authority vested in the controller}''. This basis supports the advancement of affective computing research while requiring measures such as data minimization, pseudonymization, and ethical review to protect participant interests.

Explicit consent would typically be required for special category emotional data, with higher standards of specificity and clarity than those needed for standard consent. Alternative legal bases for unique category data include substantial public interest or scientific research purposes, provided that appropriate safeguards are in place. However, these exceptions are interpreted narrowly and are subject to additional requirements.

\subsubsection{The EU Artificial Intelligence Act and Emotional Recognition}\label{the-eu-artificial-intelligence-act-and-emotional-recognition}

The European Union Artificial Intelligence Act represents a significant advancement in regulating AI systems, with specific provisions relevant to emotional analysis and affective computing. The AI Act adopts a risk-based approach, categorizing AI systems based on potential harm. Systems posing unacceptable risks are prohibited entirely, including those that deploy subliminal techniques to manipulate behavior in a harmful way. High-risk systems with a significant potential impact on health, safety, or fundamental rights are subject to stringent requirements before market introduction. Limited risk systems, including emotion recognition systems, have specific transparency obligations. Systems with minimal risk are subject to limited or no regulation under the Act.

Under the AI Act, emotion recognition systems are addressed explicitly in Article 50(3), which states: ``\textit{Deployers of an emotion recognition system or a biometric categorisation system shall inform the natural persons exposed thereto of the operation of the system, and shall process the personal data in accordance with Regulations (EU) 2016/679 and (EU) 2018/1725 and Directive (EU) 2016/680, as applicable. This obligation shall not apply to AI systems used for biometric categorisation and emotion recognition, which are permitted by law to detect, prevent or investigate criminal offences, subject to appropriate safeguards for the rights and freedoms of third parties, and in accordance with Union law}''. This provision establishes a clear transparency obligation for developers and deployers of emotion recognition systems, requiring them to explicitly disclose to individuals when such systems are in use. This represents a significant advancement over previous regulatory frameworks, which did not specifically address emotion recognition technologies.

If an emotion recognition system is classified as high-risk — such as when used in employment, education, or law enforcement contexts — it would be subject to additional requirements under the AI Act. These requirements include implementing a risk management system that identifies and mitigates risks throughout the system's lifecycle, as well as establishing data governance protocols. This ensures that training, validation, and testing datasets meet quality criteria, including relevance, representativeness, accuracy, and completeness. High-risk systems must also maintain comprehensive technical documentation that details the system's development, functionality, and compliance with requirements. The automatic logging of the AI system's operation must be kept for record-keeping purposes. To ensure transparency, users must have clear information about the system's capabilities, limitations, and intended purpose. Human oversight measures must be implemented to ensure humans can effectively supervise the system. Ultimately, the system must achieve the necessary accuracy, robustness, and cybersecurity levels, including resilience to errors and protection against unauthorized access.

The AI Act prohibits certain practices, particularly relevant to emotion recognition and manipulation. These include AI systems that deploy subliminal techniques beyond a person's conscious awareness to distort behavior that causes physical or psychological harm. Systems that exploit vulnerabilities of specific groups based on age, disability, or social or economic situation to materially distort behavior in a harmful way are also prohibited.
Additionally, the Act prohibits social scoring by public authorities, which involves evaluating or classifying individuals based on their social behavior or personal characteristics, potentially leading to detrimental or unfavorable treatment in unrelated social contexts. These prohibitions provide necessary guardrails against the most harmful potential applications of emotional intelligence in AI systems, particularly those that might exploit emotional vulnerabilities for manipulative purposes.

\subsubsection{Other Relevant Regulatory Frameworks}\label{other-relevant-regulatory-frameworks}

Beyond the GDPR and AI Act, several other regulatory frameworks have implications for emotional data and affective computing. Biometric privacy regulations at the state level, including the Illinois Biometric Information Privacy Act (BIPA), impose specific requirements for collecting and processing biometric identifiers and biometric information. These laws typically require informed written consent before the collection, disclosure of the purpose, and the length of term for which the data will be collected, stored, and used, as well as a publicly available written policy establishing a retention schedule and guidelines for destruction.
As emotionally intelligent AI systems increasingly rely on biometric methods for emotion recognition, compliance with these specialized privacy laws becomes increasingly relevant.

Consumer protection authorities have begun addressing the collection of emotional data in commercial contexts. 

In healthcare contexts, emotional data may be subject to specialized health information privacy laws such as the Health Insurance Portability and Accountability Act (HIPAA) in the United States. These regulations impose strict requirements on collecting, using, and disclosing protected health information, which could include emotional data when used for health-related purposes such as mental health assessment or monitoring.

Specialized protections for children's data, such as the Children's Online Privacy Protection Act (COPPA) in the United States and provisions within the GDPR (Article 8), place additional restrictions on collecting and processing data from children. These protections are particularly relevant to emotional data, given children's potentially heightened vulnerability to emotional analysis and manipulation.

\subsection{Case Study: ChatGPT-4.5 and Emotional Intelligence}\label{case-study-chatgpt-4.5-and-emotional-intelligence}

\subsubsection{Emotion Recognition in Text}\label{emotion-recognition-in-text}

OpenAI's release of ChatGPT-4.5 in late 2023 marked a significant advancement in the emotional intelligence capabilities of large language models.

ChatGPT-4.5 employs several techniques to recognize emotional content in user inputs. The model identifies the overall positive, negative, or neutral sentiment expressed in the text, providing a foundational layer of emotional awareness. Beyond basic sentiment, it categorizes text into specific emotional categories, including primary emotions (joy, sadness, anger, fear, surprise, disgust) and secondary (pride, shame, guilt, envy, etc.). The system assesses the strength or intensity of expressed emotions, distinguishing between mild annoyance and intense rage, for example. It also interprets emotional content within a conversational context, recognizing how emotions evolve throughout an interaction and how prior exchanges inform current emotional states.
Beyond explicit emotional statements, the model identifies implicit emotional cues, including sarcasm, passive aggression, excitement, and uncertainty conveyed through word choice, syntax, and stylistic elements.

These capabilities are implemented through supervised learning on annotated emotional data, such as Reinforcement Learning from Human Feedback (RLHF) in Machine Learning, which prioritizes emotionally appropriate responses, and few-shot learning that enables adaptation to individual users' emotional expression patterns.

\subsubsection{Emotional Response Generation}\label{emotional-response-generation}

ChatGPT-4.5's response generation incorporates emotional intelligence through several mechanisms. The model can calibrate its response tone to match the user's emotional state, validating and recognizing expressed emotions through empathetic mirroring. For users expressing distress, the model offers responses designed to facilitate healthy emotional processing, including validation, normalization, reframing, and resource suggestion as part of emotional regulation support. The system modulates emotional content based on conversation context, distinguishing between scenarios where emotional engagement is beneficial versus situations requiring more neutral, factual responses. Through conversation history, the model develops user-specific emotional profiles that inform future interactions, allowing for more personalized emotional support. The model also adapts its emotional responses based on cultural contexts, recognizing variations in emotional display rules, expression norms, and values across different cultural backgrounds.

\subsubsection{Regulatory Compliance and Ethical Considerations}\label{regulatory-compliance-and-ethical-considerations}

OpenAI has not implemented user interface elements that indicate real-time emotional analysis, nor are there visual cues, confidence indicators for emotional inferences, or specific disclosures on the emotional interpretability of the model’s outputs. The “About” section in the ChatGPT interface provides general information about the system’s capabilities, limitations, and safety measures. Still, it does not provide detailed guidance or transparency regarding the processing of emotional data.

OpenAI’s current privacy approach includes standard user data handling mechanisms, such as data deletion via user controls and retention policies aligned with privacy laws like the GDPR and CCPA. However, there is no publicly documented system of layered consent specifically for emotional data processing, nor is there a separate framework for specifying purposes, managing emotional profiles, or implementing age-specific emotional safeguards. The platform allows users to delete individual conversations, which may include sensitive content; however, this is not tied to a distinct emotional data processing framework.

To date, OpenAI has not disclosed using on-device emotional analysis or local processing options to minimize cloud-based data handling for emotional content. Additionally, while OpenAI incorporates differential privacy and other techniques in research and model training contexts, these techniques are not explicitly documented as being applied to emotional pattern extraction or anonymized emotional profiles.

Despite the absence of native emotional intelligence systems, the broader emergence of emotionally responsive AI, including third-party applications built on top of large language models, has sparked widespread discussions on ethics and regulation. Scholars, privacy advocates, and ethicists have raised concerns about the potential for emotionally interactive systems to manipulate users, particularly in commercial or political settings. The simulation of empathy by AI has sparked debates about deception, authenticity, and the psychological implications of human-AI interactions. Mental health professionals have warned against overreliance on AI for emotional support, citing the limitations of machine systems in terms of clinical judgment, empathy, and ethical oversight. Due to its intimate and potentially revealing nature, privacy experts argue that emotional data may warrant special protection. Additionally, labor theorists have suggested that emotionally interactive systems risk extracting emotional labor from users — data that helps refine model responsiveness — without compensation or acknowledgment.

On February 27, 2025, OpenAI released a research preview of OpenAI GPT‑4.5 (OpenAI GPT-4.5 System Card), and in the related \href{https://openai.com/index/gpt-4-5-system-card/}{communication}, among other things, it is stated:

\begin{quote}

«Early testing shows that interacting with GPT‑4.5 feels more natural. Its broader knowledge base, stronger alignment with user intent, and \textbf{\textit{improved emotional intelligence}} make it well-suited for tasks like writing, programming, and solving practical problems—with \textbf{\textit{fewer hallucinations}}.».

\end{quote}

Two fundamental aspects emerge from the communication above. First, the \textbf{improvement of emotional intelligence} is highlighted, and the term "improvement" is very significant. This is precisely to emphasize that this is not an absolute novelty, but rather an existing feature subject to further experimentation. Therefore, the ChatGPT AI system has been enhanced to improve its emotional profile, which should lead users to a better understanding of the requests made to the LLM system and the responses provided by it. Second, OpenAI also states that \textbf{hallucinations} have been \textbf{reduced} in the new 4.5 version of ChatGPT (refer to the section on LLMs for a description of this phenomenon). Hallucinations are one of the primary risks associated with LLMs, as they generate output that cannot be found in reality; often, the content does not exist and is an invention of the system.

\subsection{Ethical Frameworks and Governance for Emotional AI}\label{ethical-frameworks-and-governance-for-emotional-ai}

\subsubsection{Ethical Principles for Emotionally Intelligent AI - Autonomy and Dignity}\label{ethical-principles-for-emotionally-intelligent-ai-autonomy-and-dignity}

A comprehensive governance framework should explicitly incorporate ethical principles that address the unique characteristics of emotionally intelligent systems.

Emotionally intelligent AI should respect human autonomy and dignity by preserving individual control over emotional expression and its implications. This principle requires that systems avoid manipulative techniques that exploit emotional vulnerabilities, thereby undermining authentic agency. Respecting the authenticity of emotional experience and communication means acknowledging that emotions represent a core aspect of human identity and experience, not merely data to be extracted and optimized. Systems should provide transparent information about their emotional capabilities and limitations, enabling users to understand the nature of their interaction with emotionally responsive AI. Furthermore, meaningful opt-out options for emotional analysis must be available, allowing individuals to engage with AI systems without subjecting their emotional expressions to algorithmic processing. These provisions collectively safeguard the fundamental human dignity intimately connected to emotional authenticity and self-determination.

\subsubsection{Beneficence and Non-maleficence}\label{beneficence-and-non-maleficence}

AI systems with emotional capabilities should prioritize user well-being in emotional interactions, aiming to enhance emotional health rather than merely maximize engagement or commercial objectives. This requires careful attention to avoid exacerbating negative emotional states or mental health conditions, which could occur through inappropriate reinforcement, amplification, or exploitation of distress. Developers must recognize the limits of automated emotional support, acknowledging that AI systems cannot fully replace human connection, therapeutic expertise, or contextual understanding in addressing complex emotional needs. Appropriate safeguards for vulnerable users---including individuals experiencing acute psychological distress, those with mental health conditions, and those with limited digital literacy---must be integrated into system design and deployment. Furthermore, ethical guidelines for responding to concerning emotional content, such as expressions of self-harm or suicidal ideation, should be developed in collaboration with mental health professionals and implemented consistently across platforms. These principles ensure that emotional AI is a beneficial complement to human emotional support rather than a potentially harmful substitute.

\subsubsection{Justice and Fairness}\label{justice-and-fairness}

Ethical deployment of emotional AI requires attention to justice and fairness across multiple dimensions. At its core, this principle demands equitable performance of emotion recognition systems across diverse demographic groups. Research has repeatedly demonstrated that many current affective computing systems exhibit significant disparities in recognition accuracy when analyzing emotional expressions from individuals of different ages, genders, ethnicities, and cultural backgrounds. These disparities often stem from training data imbalances, where specific populations are overrepresented while others remain marginalized or absent. The consequences of such imbalances extend beyond mere technical performance metrics, potentially reinforcing existing social inequalities by providing superior service to privileged groups while delivering substandard experiences to historically underserved populations.

Furthermore, justice in emotional AI necessitates acknowledging and accommodating cultural variations in emotional expression and interpretation. The dominant psychological models of emotion that inform most affective computing systems emerged primarily from Western, educated, industrialized, rich, and democratic (WEIRD) populations, yet are often inappropriately generalized as universal. Cross-cultural research has revealed substantial differences across societies in emotional display rules, vocabulary, interpretation, and expression norms. For instance, the concepts of "grief" or the expression of "joy" may manifest differently in collectivist versus individualist cultures, while some emotional states may have no direct translation across cultures. Emotionally intelligent systems that fail to account for these cultural differences risk misinterpreting or misrepresenting users' emotional states based on culturally biased assumptions.

Another crucial aspect of fairness concerns the avoidance of emotional stereotyping based on demographic characteristics. Many current emotion recognition systems implicitly embed problematic assumptions, such as expectations that women will express emotions more intensely than men or that certain ethnic groups display specific emotional patterns. These stereotypes not only lead to inaccurate assessments but can also perpetuate harmful social biases. Just as human assessments of emotion can be distorted by prejudice, algorithmic systems can encode and amplify such biases at scale if not carefully designed and monitored.

Justice in emotional AI also extends to accessibility considerations.
People with neurological differences, such as those on the autism spectrum or individuals with facial mobility limitations due to conditions like Parkinson's disease, may express emotions in ways that diverge from normative patterns. Systems designed only to recognize typical emotional expressions may systematically misinterpret or fail to register the emotional states of these individuals, creating significant barriers to equitable access. Truly just emotional AI must accommodate diverse modes of emotional expression, including atypical patterns that may arise from neurological or physical differences.

The principle of fairness further demands transparent disclosure of emotional AI capabilities and limitations. Users should be informed about the system's emotional recognition functionalities, including information about its accuracy rates across different demographic groups and emotional categories. This transparency enables informed decision-making about whether and how to engage with emotionally intelligent systems, particularly for individuals from groups subject to higher error rates or misinterpretations.
Without such disclosure, users cannot meaningfully assess the system's trustworthiness or appropriateness for their needs.

Finally, justice requires meaningful access to redress mechanisms when emotional AI systems cause harm. Users must have practical, accessible means to contest inaccurate emotional assessments, particularly when these assessments lead to adverse outcomes in hiring, education, healthcare, or legal proceedings. These mechanisms should not place undue burdens on affected individuals and should include options for independent verification and correction of systemic biases rather than merely addressing individual complaints.

\subsubsection{Multi-stakeholder Governance Approach}\label{multi-stakeholder-governance-approach}

The complex ethical, legal, and social implications of Emotional AI necessitate a governance approach that engages diverse stakeholders in complementary roles. No single entity — whether governmental, corporate, or civil society —possesses the comprehensive perspective needed to address the multifaceted challenges that emotional intelligence systems present. Instead, effective governance requires coordinated action across multiple domains.

Regulatory bodies at national and international levels play essential roles in establishing baseline requirements for emotional AI systems.
These include mandatory disclosure of emotional recognition capabilities, minimum standards for accuracy and fairness across demographic groups, limitations on specific applications (such as emotion recognition in high-stakes decision-making), and enforcement mechanisms for violations. The European Union's AI Act represents a significant step in this direction, with its explicit attention to emotion recognition technologies and risk-based regulatory approach.
However, formal regulation must be supplemented by more dynamic and responsive governance mechanisms.

Industry self-regulation complements governmental oversight, enabling rapid adaptation to technological developments and emerging risks. Industry initiatives can establish technical standards for emotional data protection, ethical, emotional AI certification programs, and best practice frameworks that exceed minimum regulatory requirements. Organizations like the \href{https://standards.ieee.org/industry-connections/activities/ieee-global-initiative/}{IEEE Global Initiative 2.0 on Ethics of Autonomous and Intelligent Systems} have begun developing such standards, though implementation remains voluntary and inconsistent across the sector. More robust industry governance would include independent assessment mechanisms, transparency in algorithmic impact assessments, and sanctions for non-compliance with established standards.

Civil society organizations contribute essential perspectives from communities that may be affected by emotional AI, providing critical scrutiny of both regulatory and industry approaches. These organizations advocate for marginalized groups whose emotional expressions may be misinterpreted or whose vulnerabilities might be exploited by emotionally intelligent systems. They also conduct independent research on social and psychological impacts, develop conceptual frameworks for understanding emotional data rights, and educate the public about interactions with emotionally intelligent systems. The involvement of civil society helps ensure that governance frameworks address not only technical performance metrics but also broader societal implications.

Academic institutions offer crucial theoretical foundations and empirical evidence to inform governance approaches. Interdisciplinary research spanning computer science, psychology, ethics, law, and social sciences provides insights into the technical limitations, psychological effects, and societal impacts of emotional AI. Academic work can identify emerging risks before they manifest at scale, evaluate the effectiveness of various governance mechanisms, and propose innovative approaches to strike a balance between innovation and protecting fundamental rights and values.

User engagement represents the most frequently overlooked but essential component of effective governance. Users of emotional AI systems should actively define acceptable practices, identify concerns, and evaluate proposed safeguards. This engagement should extend beyond token consultation to meaningful participation in system design, policy development, and impact assessment. User perspectives are particularly valuable in identifying the contextual factors that influence emotional analysis and response appropriateness in different situations.

\subsubsection{Future Directions and Recommendations}\label{future-directions-and-recommendations}

As affective computing advances and emotional intelligence becomes increasingly integrated into AI systems, several key research, policy, and practice directions emerge.

First, research into emotional diversity must be prioritized to address current limitations in recognizing and responding to human emotional experiences. This includes expanded studies of emotional expression across cultures, investigation of emotional communication in neurodiverse populations, and exploration of complex, mixed, and culturally specific emotional states that may not fit neatly into dominant psychological models. Such research should employ participatory methods that engage diverse communities as collaborators rather than merely subjects of study.

Second, privacy-preserving techniques for emotion recognition warrant significant investment. Current approaches often require the extensive collection and centralized processing of sensitive, emotional data, creating substantial privacy risks. Techniques like federated learning, differential privacy, and on-device processing could enable emotional intelligence while minimizing data exposure. Similarly, systems should be designed to forget emotional data after use rather than accumulating increasingly comprehensive emotional profiles of individuals over time.

Third, context-sensitive governance frameworks must be developed to address the varying implications of emotional AI across different domains. The appropriate standards, safeguards, and limitations for emotional analysis in healthcare settings differ substantially from those needed in educational contexts, workplace environments, or consumer applications. Domain-specific guidelines should be co-created with relevant stakeholders, including professionals working in these fields and the individuals whose emotional data might be processed.

Fourth, interdisciplinary education and training programs are necessary to develop professionals who can navigate the technical, ethical, legal, and social dimensions of emotional AI. Current educational pathways typically separate these domains, resulting in computer scientists with a limited understanding of emotional psychology, psychologists with minimal technical knowledge of AI systems, and legal scholars unfamiliar with both. Integrated approaches to education would better prepare the next generation of researchers, developers, and policymakers to address the complex challenges of emotionally intelligent systems.

Fifth, ongoing monitoring and evaluation mechanisms should be established to assess the impacts of emotional AI systems after deployment. These mechanisms should track technical performance metrics, psychological effects on users, social consequences for different communities, and evolving public attitudes toward emotional AI. Findings from such assessments should inform iterative improvements to technical systems and governance frameworks.

Ultimately, international cooperation on the governance of emotional AI should be strengthened to prevent regulatory fragmentation and establish shared principles across jurisdictions. While cultural differences may necessitate some variation in specific standards and approaches, core values such as respect for human dignity, protection of vulnerable populations, and preservation of emotional autonomy can provide common ground for global governance efforts.

\section{Conclusion}\label{conclusion}

Integrating emotional intelligence into AI systems represents a transformative opportunity and a significant challenge for technology governance. Affective computing enables more natural and responsive human-computer interaction, opening new possibilities for healthcare, education, and personal well-being applications. However, as machines gain an increased capacity to recognize, interpret, and respond to human emotions, we must carefully consider the implications for privacy, autonomy, fairness, and the fundamental nature of emotional experience.

This paper examines the technical foundations of affective computing, which involves transforming human emotions into structured data, and the emerging regulatory frameworks governing emotional AI. Through analysis of current approaches and the case study of ChatGPT-4.5, we have identified key challenges in ensuring the responsible development and deployment of emotionally intelligent systems. These challenges cannot be addressed solely through technical solutions or resolved by regulatory intervention alone. Instead, they require a multidimensional approach integrating technical innovation, ethical reflection, legal frameworks, and inclusive stakeholder engagement.

The path forward demands balancing seemingly competing values: advancing emotional AI capabilities while protecting privacy, enabling personalization while preventing manipulation, recognizing universal aspects of emotion while respecting cultural diversity, and maintaining human connection while incorporating technological mediation. Navigating these tensions requires ongoing dialogue across disciplines and sectors, with a particular focus on voices that have been historically marginalized in technology governance.

As affective computing evolves, our choices about data practices, system design, regulatory approaches, and governance structures will shape the future relationship between humans and emotionally intelligent machines. By emphasizing human dignity, autonomy, and well-being as guiding principles in this development, we can harness the potential of emotional AI while safeguarding the richness and authenticity of human emotional experience in the digital age.

\section{References}\label{references}

Barrett, L. F. (2017). How emotions are made: The secret life of the brain. Houghton Mifflin Harcourt.

Busso, C., Bulut, M., Lee, C. C., Kazemzadeh, A., Mower, E., Kim, S., Chang, J. N., Lee, S., \& Narayanan, S. S. (2008). IEMOCAP: Interactive emotional dyadic motion capture database. Language Resources and Evaluation, 42(4), 335-359.

California Privacy Rights Act. (2020). California Civil Code § 1798.100-1798.199.100.

Devlin, J., Chang, M. W., Lee, K., \& Toutanova, K. (2019). BERT: Pre-training of deep bidirectional transformers for language understanding. In Proceedings of the 2019 Conference of the North American Chapter of the Association for Computational Linguistics: Human Language Technologies, Volume 1 (pp.~4171-4186).
Ekman, P. (1992). An argument for basic emotions. Cognition \& Emotion,
6(3-4), 169-200.

Ekman, P., \& Friesen, W. V. (1982). Felt, false, and miserable smiles. Journal of Nonverbal Behavior, 6(4), 238-252.

REGULATION (EU) 2024/1689 OF THE EUROPEAN PARLIAMENT AND OF THE COUNCIL of 13 June 2024 laying down harmonised rules on artificial intelligence and amending Regulations (EC) No 300/2008, (EU) No 167/2013, (EU) No 168/2013, (EU) 2018/858, (EU) 2018/1139 and (EU) 2019/2144 and Directives 2014/90/EU, (EU) 2016/797 and (EU) 2020/1828 (Artificial Intelligence Act).

Fabian Benitez-Quiroz, C., Srinivasan, R., \& Martinez, A. M. (2016).
EmotioNet: An accurate, real-time algorithm for the automatic annotation of a million facial expressions in the wild. In Proceedings of the IEEE 
Conference on Computer Vision and Pattern Recognition (pp.~5562-5570).

Felbo, B., Mislove, A., Søgaard, A., Rahwan, I., \& Lehmann, S. (2017).
Using millions of emoji occurrences to learn any-domain representations
for detecting sentiment, emotion and sarcasm. In Proceedings of the 2017
Conference on Empirical Methods in Natural Language Processing
(pp.~1615-1625).

Goodfellow, I. J., Erhan, D., Carrier, P. L., Courville, A., Mirza, M.,
Hamner, B., Cukierski, W., Tang, Y., Thaler, D., Lee, D. H., Zhou, Y.,
Ramaiah, C., Feng, F., Li, R., Wang, X., Athanasakis, D., Shawe-Taylor,
J., Milakov, M., Park, J., \ldots{} \& Bengio, Y. (2013). Challenges in
representation learning: A report on three machine learning contests.
Neural Networks, 64, 59-63.

He, K., Zhang, X., Ren, S., \& Sun, J. (2016). Deep residual learning
for image recognition. In Proceedings of the IEEE Conference on Computer
Vision and Pattern Recognition (pp.~770-778).

Hochreiter, S., \& Schmidhuber, J. (1997). Long short-term memory.
Neural Computation, 9(8), 1735-1780.

LeCun, Y., Bengio, Y., \& Hinton, G. (2015). Deep learning. Nature,
521(7553), 436-444.

McKeown, G., Valstar, M., Cowie, R., Pantic, M., \& Schroder, M. (2012).
The SEMAINE database: Annotated multimodal records of emotionally
colored conversations between a person and a limited agent. IEEE
Transactions on Affective Computing, 3(1), 5-17.

Picard, R. W. (1997). Affective computing. MIT Press.

Robinson, M. D., \& Clore, G. L. (2002). Belief and feeling: Evidence
for an accessibility model of emotional self-report. Psychological
Bulletin, 128(6), 934-960.

Tan, M., \& Le, Q. (2019). EfficientNet: Rethinking model scaling for
convolutional neural networks. In Proceedings of the 36th International
Conference on Machine Learning (pp.~6105-6114).

\end{document}